\documentclass{caosp}

\usepackage{graphicx}

\articleNo{123}
\pubyear{2005}
\volume{35}
\volnumber{3}
\firstpage{1}
\received{May 1, 2005}
\accepted{August 28, 2005}

\begin{document}

\hauthor{O. Kochukhov et al.}

\title{A search of very low amplitude magnetoacoustic pulsations with HARPS}


\author{O. Kochukhov \inst{1} \and 
        T. Ryabchikova \inst{2,} \inst{3} \and
        S. Bagnulo \inst{4} \and
        G. Lo Curto \inst{5}}

\institute{
Department of Astronomy and Space Physics, 
Uppsala University, Box 515, SE-751 20 Uppsala, Sweden
\and
Institute for Astronomy, Russian Academy of Sciences, 
Pyatnitskaya 48, 119017 Moscow, Russia
\and 
Institute for Astronomy, University of Vienna, 
T\"urkenschanzstra{\ss}e 17, A-1180 Vienna, Austria
\and
Armagh Observatory, College Hill, Armagh, BT61 9DG, \\ Northern 
Ireland, UK
\and
European Southern Observatory, Alonso de Cordova 3107, 
Vitacura, Santiago, Casilla 19001 Santiago 19, Chile
%
}

\date{December 1, 2007}

\maketitle

\begin{abstract}
We have obtained time-resolved spectroscopic observations for a sample of 10 cool Ap stars using
the ultra-stable spectrograph HARPS at the ESO 3.6-m telescope. The aim of our study was to search
for low-amplitude oscillations in Ap stars with no or inconclusive evidence of pulsational
variability. Here we report initial results of our investigation.
We confirm the presence of $\approx$\,16-min period pulsations in $\beta$~CrB (HD~137909) and
demonstrate multiperiodic character of oscillations in this star. Furthermore, we discovered very
low amplitude 9-min pulsations in HD~75445 -- an object spectroscopically very similar to known roAp
stars.
\keywords{stars: atmospheres -- stars: chemically peculiar -- 
stars: oscillations -- stars: individual: HD\,75445, HD\,137909}
\end{abstract}

\section{Introduction}

Many recent time-resolved spectroscopic studies of the rapidly oscillating Ap (roAp) stars (e.g.,
Kurtz et al. \cite{kc06}, Ryabchikova et al. \cite{tr07}) have focused on the known, high-amplitude
pulsators and did not address the fundamental question of what distinguishes pulsating and
non-oscillating Ap (noAp) stars. Ryabchikova et al. (\cite{tr04}) suggested that there might exist
no real dichotomy between roAp and noAp stars, and weak pulsational variability is present in all
cool Ap stars.  With the aim to test this hypothesis and to obtain the first complete picture of the
incidence of pulsations in Ap stars, we used the ultra-stable spectrograph HARPS at ESO to obtain
time-resolved spectra for a sample of 10 noAp and weakly oscillating Ap stars. Here we summarize
preliminary results of the analysis of HARPS spectra of HD\,75445 and HD\,137909 ($\beta$~CrB).

\begin{figure}[!t]
\includegraphics[width=6cm]{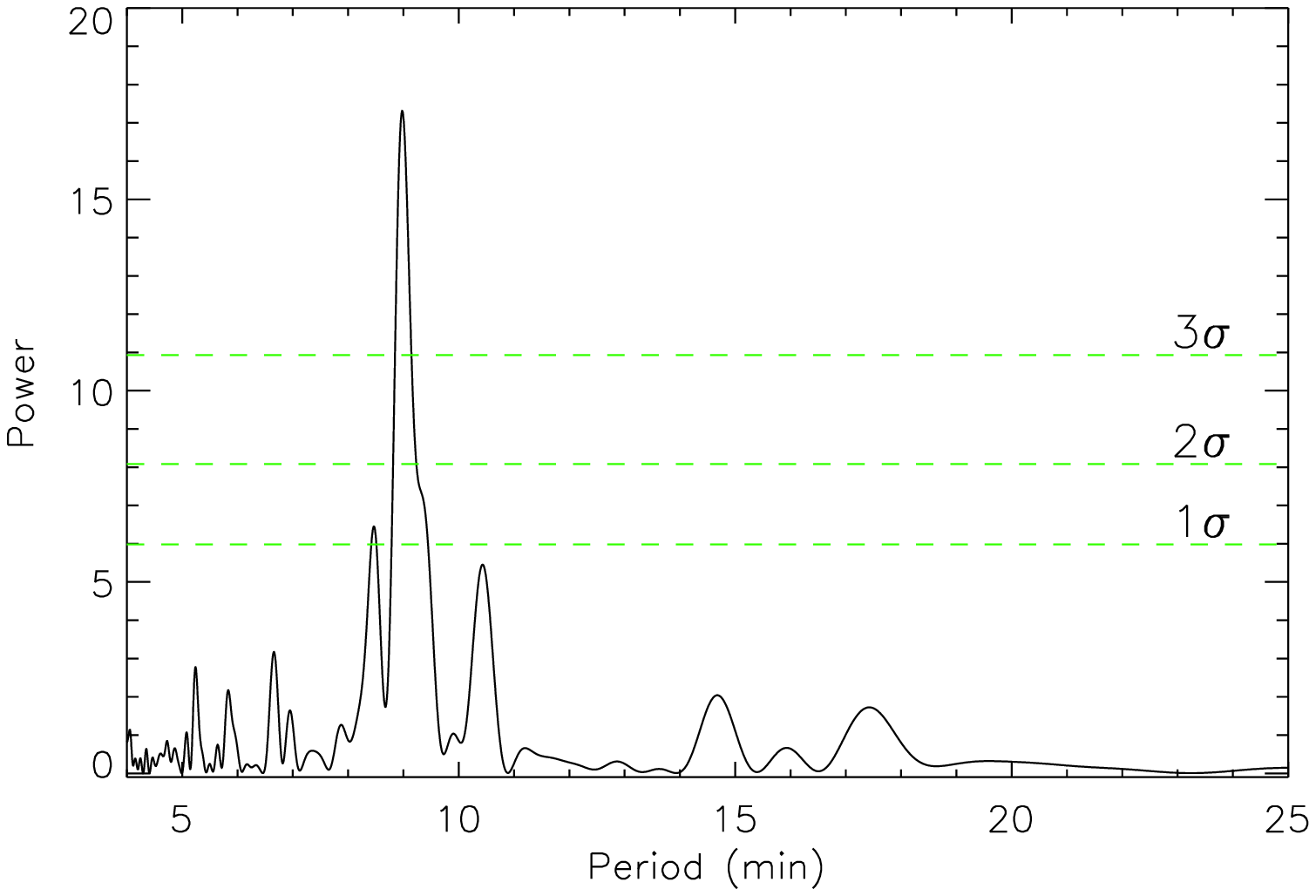}
\includegraphics[width=6.0cm]{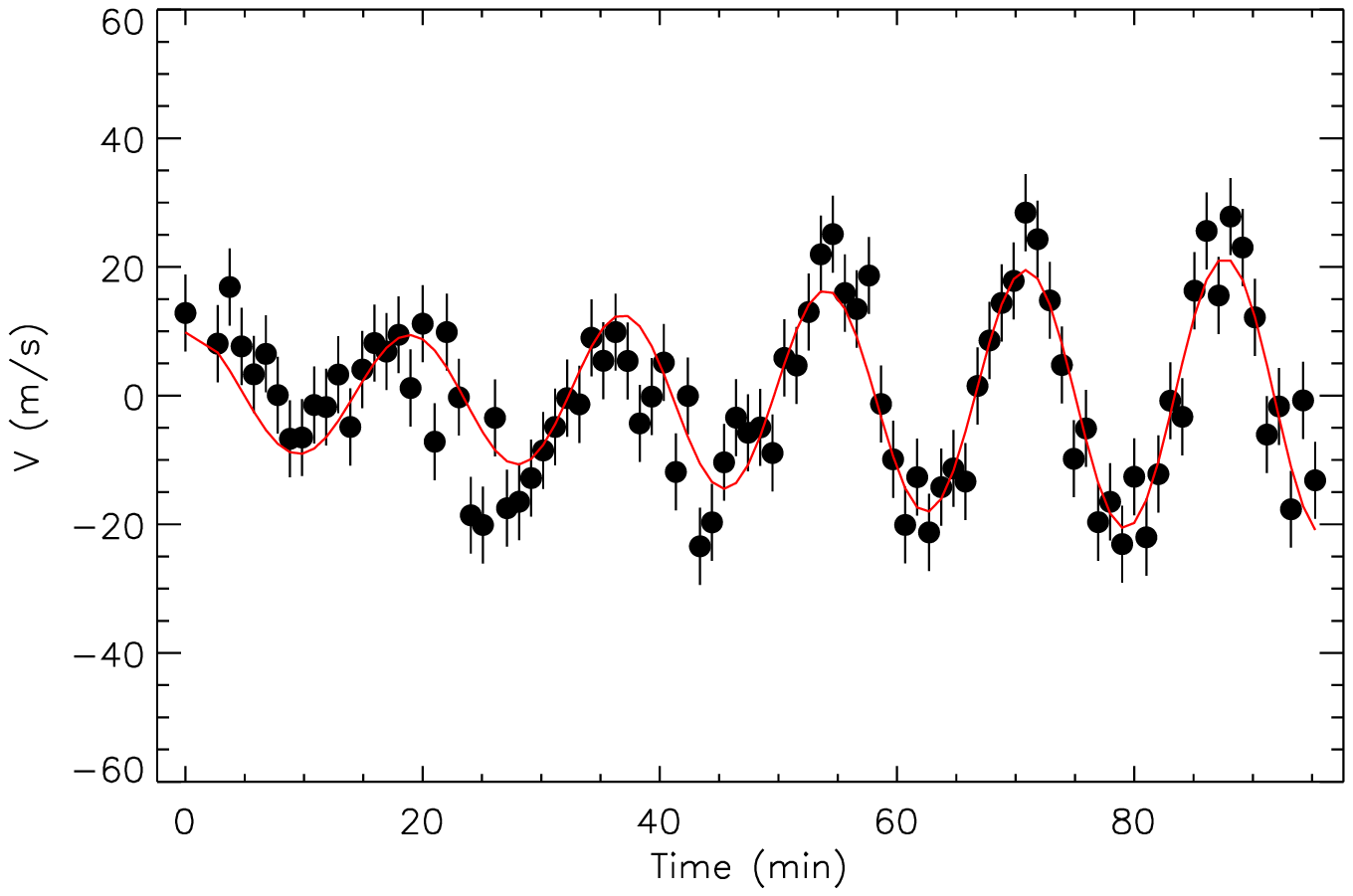}
\caption{{\it Left panel:} Periodogram showing discovery of 9-min pulsational signal in the 
combined RVs of 29 Nd~{\sc iii} lines in HD\,75445.
{\it Right panel:} Mean RV variation of 40 Ce~{\sc ii}
lines in $\beta$~CrB. Multiperiodic character of oscillation
is clearly evident. 
}
\label{fig1}
\end{figure}

\section{HD 75445}

Ryabchikova et al. (\cite{tr04}) performed an abundance analysis of HD\,75445, finding this star to
be a spectroscopic twin of the well-known roAp star $\gamma$~Equ. 
%
We have obtained a set of 120 HARPS spectra of HD\,75445. Pulsational analysis of Nd~{\sc iii} lines
reveals oscillations with 8.98~min period and an amplitude of $20\pm3$ m\,s$^{-1}$.
Thus, HD\,75445 is a new roAp star, and the first one discovered with HARPS.

\section{$\beta$~CrB}

Kochukhov et al. (\cite{ok02}) found a tentative evidence of the 11.5-min RV variation  in
$\beta$~CrB. This period was not confirmed by Hatzes \& Mkrtichian (\cite{hm04}), who reported
definite 16.2-min oscillation. 
%
Using HARPS we obtained 93 high-precision observations of $\beta$~CrB. The RV analysis of 
Ce~{\sc ii} lines 
reveals pulsational variability close to the 
previously reported 16.2-min period. Moreover, our time-series shows strong amplitude modulation
(Fig.~\ref{fig1}), indicating the presence of at least two frequencies. A least-squares fit yields
periods $17.2\pm0.2$ and $15.7\pm0.5$ min with amplitudes $15.4\pm2$ and $6.3\pm2$ m\,s$^{-1}$,
respectively.


\begin{thebibliography}{}
\bibitem[2004]{hm04}
Hatzes, A. P., \& Mkrtichian, D. E. 2004, \mnras, 351, 663
\bibitem[2002]{ok02}
Kochukhov, O., Landstreet, J. D., Ryabchikova, T., Weiss, W. W., \& Kupka, F. 2002, \mnras, 337, L1
\bibitem[2006]{kc06}
Kurtz, D. W., Elkin, V. G., \& Mathys, G. 2006, \mnras, 370, 1274
\bibitem[2004]{tr04}
Ryabchikova, T., Nesvacil, N., Weiss, W. W., Kochukhov, O., \& St\"utz, C. 2004, \aaa, 423, 705
\bibitem[2007]{tr07}
Ryabchikova, T., Sachkov, M., Kochukhov, O., \& Lyashko, D. 2007, \aaa, 473, 907
\end{thebibliography}
\end{document}